# Self-citation as an Impact-Reinforcing Mechanism in the Science System

**Anthony F. J. van Raan**
Center for Science and Technology Studies
Leiden University
Wassenaarseweg 52
P.O. Box 9555
2300 RB Leiden, Netherlands

***Abstract***

*In previous papers it was demonstrated that lower performance groups have a larger size-dependent cumulative advantage for receiving citations than top-performance groups. Furthermore, regardless of performance, larger groups have less not-cited publications. Particularly for the lower performance groups the fraction of not-cited publications decreases considerably with size. These phenomena can be explained with a model in which self-citation acts as a promotion mechanism for external citations. In this article we show that for self-citations similar size-dependent scaling rules apply as for citations but generally the power law exponents are higher for self-citations as compared to citations. We also find that the fraction of self-citations is smaller for the higher performance groups and this fraction decreases more rapidly with increasing journal impact than for lower performance groups. An interesting novel finding is that the variance in the correlation of the number of self-citations with size is considerably less than the variance for external citations. This is a clear indication that size is a stronger determinant for self-citations than for external citations. Both higher and particularly lower performance groups have a size-dependent cumulative advantage for self-citations, but for the higher performance groups only in the lower impact journals and in fields with low citation density.*

## 1. Introduction

There is a long history of the construction of bibliometric indicators (van Raan 2004) and there is much recent work on the use of publication and citation data in the study of author-, publication- and citation-networks in science (Albert & Barabási 2002; Dorogovtsev & Mendes 2002; Leicht, Clarkson, Shedden, & Newman 2007). But there is little work on the mutual coherence of bibliometric indicators and their statistical properties in the context of science as an interconnected system. Building on previous published work (van Raan 2006a,b; 2008a,b) the author continues in this paper the exploration of these interdependencies of the science system as a landscape characterized by field-specific citation densities. The focus in this article on the role of self-citations as a impact-reinforcing mechanism.

In previous work (van Raan 2006b) the author distinguished between top-performance and lower performance research groups in the analysis of statistical properties of bibliometric characteristics of research groups. The crucial finding was that particularly the lower performance groups have a size-dependent (size of a research group in terms of number of publications) *cumulative* advantage[1] advantage for receiving citations. Regardless of performance, larger groups have less not-cited publications. By distinguishing again between top- and lower-performance groups, it was also found that particularly for the lower performance groups the fraction of not-cited publications decreases considerably with size. The observations at the large scale level suggest two mechanisms at the local interaction level: Mechanism A concerns the not-cited publications, Mechanism B concerns the citation density of the field. In the science system both mechanisms are at work. The author discussed Mechanism B -the influence of field-specific citation characteristics on the impact of research groups- in a recent paper (van Raan 2008b). In this paper the focus is on Mechanism A.

In this mechanism self-citations play a crucial role. Self-citations as a phenomenon in the scientific communication system are a regular topic of discussion. Most of these studies concern the macro-level, e.g., countries, or fields of science and often they address the question at what level of aggregation do self-citations seriously affect the validity of bibliometric indicators for evaluation purposes. Important issues are the ageing of self-citations, the interdependencies of self- and non-self citations, the relation between the share of self-citations with other citation-based indicators, the influence of international collaboration and of co-authorships on self-citation practices, and the role of self-citations as a 'normal' aspect of the scientific communication process (van Raan 1998; Aksnes 2003; Glänzel, Thijs and Schlemmer 2004; Glänzel and Thijs 2004a,b; Thijs and Glänzel 2006; Glänzel, Debackere, Thijs and Schubert 2006). There are, however, to the best or our knowledge, no or hardly any extensive analyses of the role and statistical properties of self-citations at the level of research groups. With this paper the author intends to put an end to this undesirable situation, because the research group is the most important working floor entity in science. The author goes a step further than the view that self-citations are a normal aspect of scientific communication, and shows that self-citations are an important driving force in strengthening the impact of the work of a research group. This idea is related to studies on the use of self-citations as a strategy to make your own earlier scientific work visible (Lawani 1982; White 2001; Aksnes 2003; Fowler and Aksnes 2007). In this article however the author does not examine whether scientists deliberately use self-citations for the visibility strategy. We show that self-citations, for whatever reason given, will, in a statistically sufficiently large entity, work as impact-reinforcing elements.

The idea behind mechanism A is that advantage by size works by a process in which the number of not-cited publications is diminished, and that this mechanism is particularly effective for the lower performance groups. Thus, the larger the number of publications in a group, the more those publications are 'promoted' which otherwise would have remained uncited. In other words, size reinforces an internal promotion mechanism. Most probably this works by initial citation of these 'stay behind' publications in other more cited publications of the group. Then authors in other groups are stimulated to take notice of these stay behind publications and they eventually decide to cite them. Consequently, the mechanism starts with within-group self-citation, and subsequently spreads. It is obvious that particularly the lower performance groups will benefit from this mechanism. Top-performance groups do not 'need' the internal promotion mechanism to the same extent as low performance groups. This explains why they show less or even no cumulative advantage by size. Therefore, the group is a crucial entity, it is not 'just a set of publications' (as it is more or less the case for journals). The group represents the social structure in which the promotion mechanism by self-citations can work.

[1] 'Cumulative advantage' means that the dependent variable (for instance, number of citations of a group, ***C***) scales in a disproportional, non-linear way (in this case: power law) with the independent variable (for instance, in the present study the 'size' of a research group, in terms of number of publications, ***P***). Thus, larger groups (in terms of ***P***) do not just receive more citations (as can be expected), but they do so increasingly more 'advantageously': groups that are twice as large as other groups receive, for instance 2.4 times more citations. For a detailed discussion the author refers to a previous paper (van Raan 2006b). For a general discussion of cumulative advantage in science the author refers to Merton (1968, 1988) and Price (1976).

Obtaining reliable data at the research group level is however by far a trivial matter. Data on the level of the individual scientists, institutions, and research fields are externally available (e.g., author names, addresses, journals, field classifications, etc.). But this is not the case at the level of research groups. The only possibility to study bibliometric characteristics of research groups with 'external data' would be to use the address information within the main organization, for instance 'Department of Biochemistry' of a specific university. However, the delineation of departments or university groups through externally available data such as the address information in international literature databases is very problematic (van Raan 2005). Furthermore, the external data has to be combined carefully with 'internally stored' data (such as personnel belonging to specific groups). These data are only available from the institutions that are the target of the analysis. As indicated above, the data used in this study are the results of evaluation studies and are therefore based on strict verification procedures in close collaboration with the evaluated groups.

The structure of this paper is as follows. In Section 2 the data material, the application of the method and the calculation of the indicators are discussed. In Section 3 the results of the data analysis are presented and in Section 4 we discuss the main outcomes of this study in the framework of the Mechanism A model.

## 2. Data, Indicators, Citation-Density Landscape

The data material is based on a large set of publications (as far as published in journals covered by the Citation Index, 'CI publications'[2]) of all academic chemistry research in a country (Netherlands) for a 10-years period (1991-2000). This material is quite unique. To our knowledge no such compilations of very accurately verified publication sets on a large scale are used for statistical analysis of the characteristics of the indicators at the research group level. The (CI-) publications were collected as part of a large evaluation study conducted by the Association of Universities in the Netherlands. For a detailed discussion of the evaluation procedure and the results the author refers to the evaluation report (VSNU 2002). In the framework of this evaluation study, an extensive bibliometric analysis was performed to support the evaluation work by an international peer committee (van Leeuwen, Visser, Moed, & Nederhof 2002). In total, the analysis involves 700 senior researchers and covers about 18,000 publications and 175,000 citations (excluding self-citations) of 157 chemistry groups at ten universities.

[2] Thomson Scientific, the former Institute for Scientific Information (ISI) in Philadelphia, is the producer and publisher of the Web of Science that covers the Science Citation Index (-extended), the Social Science Citation Index and the Arts & Humanities Citation Index. Throughout this paper the author uses the term 'CI' (Citation Index) for the above set of databases.

The indicators are calculated on the basis of a total time-period analysis. This means that publications are counted for the entire 10-year period (1991-2000) and citations are counted up to and including 2000 (e.g., for publications from 1991, citations are counted from 1991 to 2000; for publications from 2000, citations are counted only in 2000). CWTS standard bibliometric indicators were applied. Here only 'external' citations, i.e., citations corrected for self-citations[3], are taken into account. These standard bibliometric indicators with a short description are presented in the text box here below. For a detailed discussion see Van Raan (1996, 2004).

**Standard Bibliometric Indicators:**

- Number of publications ***P*** in CI-covered journals of a research group in the specified period;
- Number of citations ***C*** received by ***P*** during the specified period without self-citations; including self-citations: $C_i$. Thus, number of self-citations $C_s = C_i - C$ and the relative amount (fraction) of self-citations is $C_s/C_i$;
- Average number of citations per publication, without self-citations (***CPP***);
- Percentage of publications not cited (in the specified period) ***Pnc***;
- Journal-based worldwide average impact as an international reference level for a research group (***JCS***, journal citation score, which is our journal impact indicator), without self-citations (on a world-wide scale!); in the case of more than one journal the average ***JCSm*** is used; for the calculation of ***JCSm*** the same publication and citation counting procedure, time windows, and article types are used as in the case of ***CPP***;
- Field-based[4] worldwide average impact as an international reference level for a research group (***FCS***, field citation score), without self-citations (on a world-wide scale!); in the case of more than one field (as almost always) the average ***FCSm*** is used; for the calculation of ***FCSm*** the same publication and citation counting procedure, time windows, and article types are used as in the case of ***CPP***; the author refers in this article to the ***FCSm*** indicator as the 'field-specific citation density';
- Comparison of the ***CPP*** of a research group with the world-wide average based on ***JCSm*** as a standard, without self-citations, indicator ***CPP/JCSm***;
- Comparison of the ***CPP*** of a research group with the world-wide average based on ***FCSm*** as a standard, without self-citations, indicator ***CPP/FCSm***;
- Ratio ***JCSm/FCSm*** is the relative, field-normalized journal impact indicator.

Table 1 shows as an example the results of the bibliometric analysis for the most important indicators for all 12 chemistry research groups of one of the ten universities (University A, groups A-01 to A-12). This table also shows that the indicator calculations allow a statistical analysis of these indicators. The internationally standardized (field-normalized) impact indicator ***CPP/FCSm*** is regarded as the 'crown' indicator. This indicator enables to observe whether the performance of a research group is significantly far below (indicator value < 0.5), below (0.5 - 0.8), around (0.8 - 1.2), above (1.2 – 1.5), or far above (>1.5) the international (western world dominated) impact standard of the field.

***Table 1****: Example of the results of the bibliometric analysis*

| *Research group* | *P* | *C* | *CPP* | *JCSm* | *FCSm* | *CPP/JCSm* | *CPP/FCSm* | *JCSm/FCSm* | $C_s/C_i$ |
|---|---|---|---|---|---|---|---|---|---|
| | | | | | | | | | |
| A-01 | 92 | 554 | 6.02 | 5.76 | 4.33 | 1.05 | 1.39 | 1.33 | 0.23 |
| A-02 | 69 | 536 | 7.77 | 5.12 | 2.98 | 1.52 | 2.61 | 1.72 | 0.16 |
| A-03 | 129 | 3,780 | 29.30 | 17.20 | 11.86 | 1.70 | 2.47 | 1.45 | 0.16 |

[3] A citation is a self-citation if any of the authors of the citing paper is also an author of the cited paper.
[4] The definition of fields based on a classification of scientific journals into *categories* developed by Thomson Scientific/ISI is used. Although this classification is not perfect, it provides a clear and 'fixed' consistent field definition suitable for automated procedures within our data-system.

| | | | | | | | | | |
|---|---|---|---|---|---|---|---|---|---|
| A-04 | 80 | 725 | 9.06 | 8.06 | 6.25 | 1.12 | 1.45 | 1.29 | 0.27 |
| A-05 | 188 | 1,488 | 7.91 | 8.76 | 5.31 | 0.90 | 1.49 | 1.65 | 0.30 |
| A-06 | 52 | 424 | 8.15 | 6.27 | 3.56 | 1.30 | 2.29 | 1.76 | 0.30 |
| A-07 | 52 | 362 | 6.96 | 4.51 | 5.01 | 1.54 | 1.39 | 0.90 | 0.16 |
| A-08 | 171 | 1,646 | 9.63 | 6.45 | 4.36 | 1.49 | 2.21 | 1.48 | 0.23 |
| A-09 | 132 | 2,581 | 19.55 | 15.22 | 11.71 | 1.28 | 1.67 | 1.30 | 0.25 |
| A-10 | 119 | 2,815 | 23.66 | 22.23 | 14.25 | 1.06 | 1.66 | 1.56 | 0.17 |
| A-11 | 141 | 1,630 | 11.56 | 17.83 | 12.30 | 0.65 | 0.94 | 1.45 | 0.29 |
| A-12 | 102 | 1,025 | 10.05 | 10.48 | 7.18 | 0.96 | 1.40 | 1.46 | 0.34 |

Particularly with a ***CPP/FCSm*** value above 1.5, groups can be considered as scientifically strong. A value above 2 indicates a very strong group and groups with values above 3 can generally be considered as excellent and comparable to top-groups at the best US universities (van Raan 2004). The ***CPP/FCSm*** indicator generally correlates well with the quality judgment of the peers (van Raan 2006a, b). Studies of large-scale evaluation procedures in which empirical material is available with data on both peer judgment as well as bibliometric indicators are quite rare. For notable exceptions, see Rinia, van Leeuwen, van Vuren, & van Raan (1998, 2001).

In Table 1 large differences in the ***FCSm*** values for the various research groups can be observed. This clearly illustrates that research groups even within one discipline (in this case chemistry) may work in fields with a high or a low field citation density. Generally high field-specific citation densities are found in basic research and low field-specific citation densities in applied research. For instance, research group A-02 is active in an applied field, catalysis research, and this group is characterized by a low field-specific citation density (***FCSm*** = 2.98). Group A-10 focuses on medicine-related basic research on human proteins, this group has a typical high field-specific citation density: ***FCSm*** = 14.25. For the total set of research groups it is found that ***FCSm*** values differ a factor of about 20, so more than an order of magnitude. Thus, these findings show that the idea of science as large collection of research groups positioned in a 'citation density landscape' makes sense.

In the lower part of the landscape a group with a low ***FCSm*** is indicated, this group however publishes in the better journals of the field, which means that ***JCSm*** > ***FCSm***, and within these top-journals the group performs very well so that ***CPP*** > ***JCSm***. In Table 1 research group A-02 is an example of this situation.

## 3. Results and discussion

### *3.1 Self-citations in relation to the basic indicators*

First we determine two overall correlations: the total number of external ('non-self') citations of research groups (***C***) as a function of size in terms of the total number of publications (***P***) of these groups, and a similar correlation for the self-citations ($C_S$). The results are shown in Figs. 1 and 2, respectively. Two striking observations can be made. First, the number of self-citations as a function of size has a larger power law exponent than in the case of external citations:

$C_S \sim P^{+1.33}$ as compared to $C \sim P^{+1.25}$ .

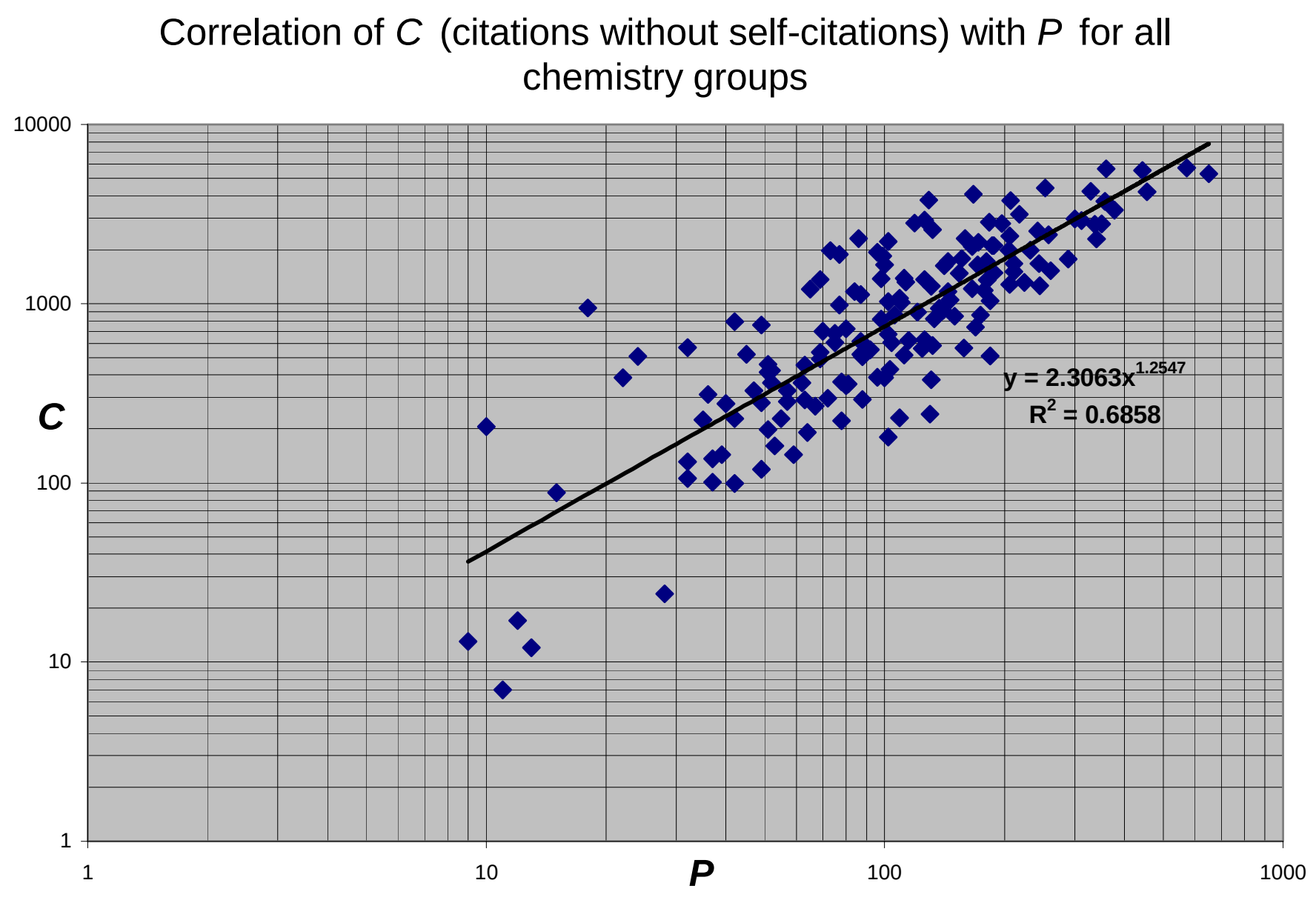

***Figure 1:*** *Correlation of the number (external) citations (**C**) received per chemistry group with the number of publications (**P**).*

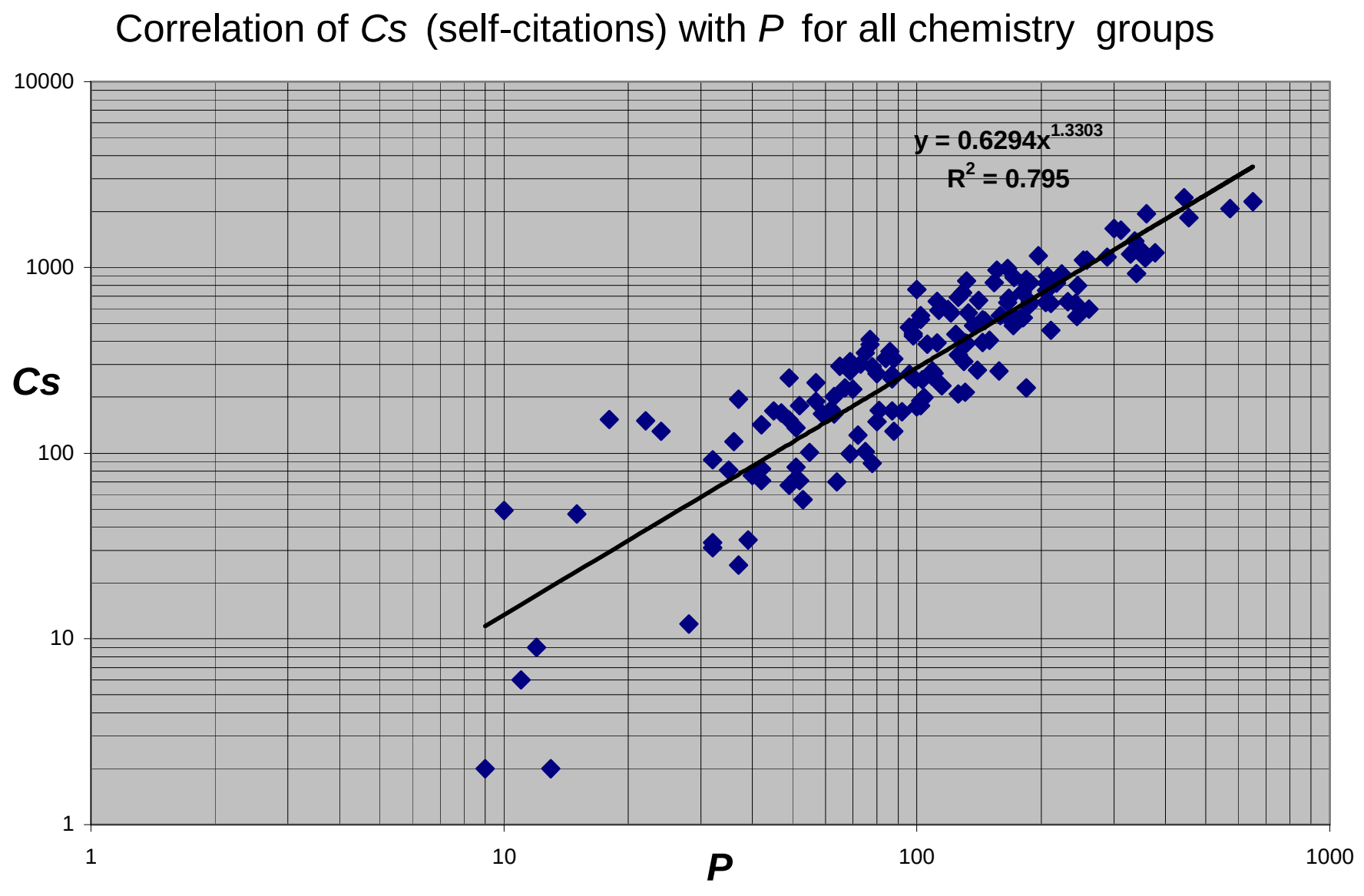

***Figure 2:*** *Correlation of the number self-citations (**$C_S$**) received per chemistry group with the number of publications (**P**).*

From the above correlations of ***C*** and ***Cs*** with ***P*** follows ***C*** ~ ***Cs***$^{+0.94}$ which is nicely confirmed by Fig. 3.

A second important observation is that the *variance in the correlation* of the number of self-citations with size is considerably less than the variance for external citations. For external citations the ‘bandwidth’ of the variance at ***P*** =100 covers a factor of around 20, for self-citations this bandwidth covers a factor of around 8.

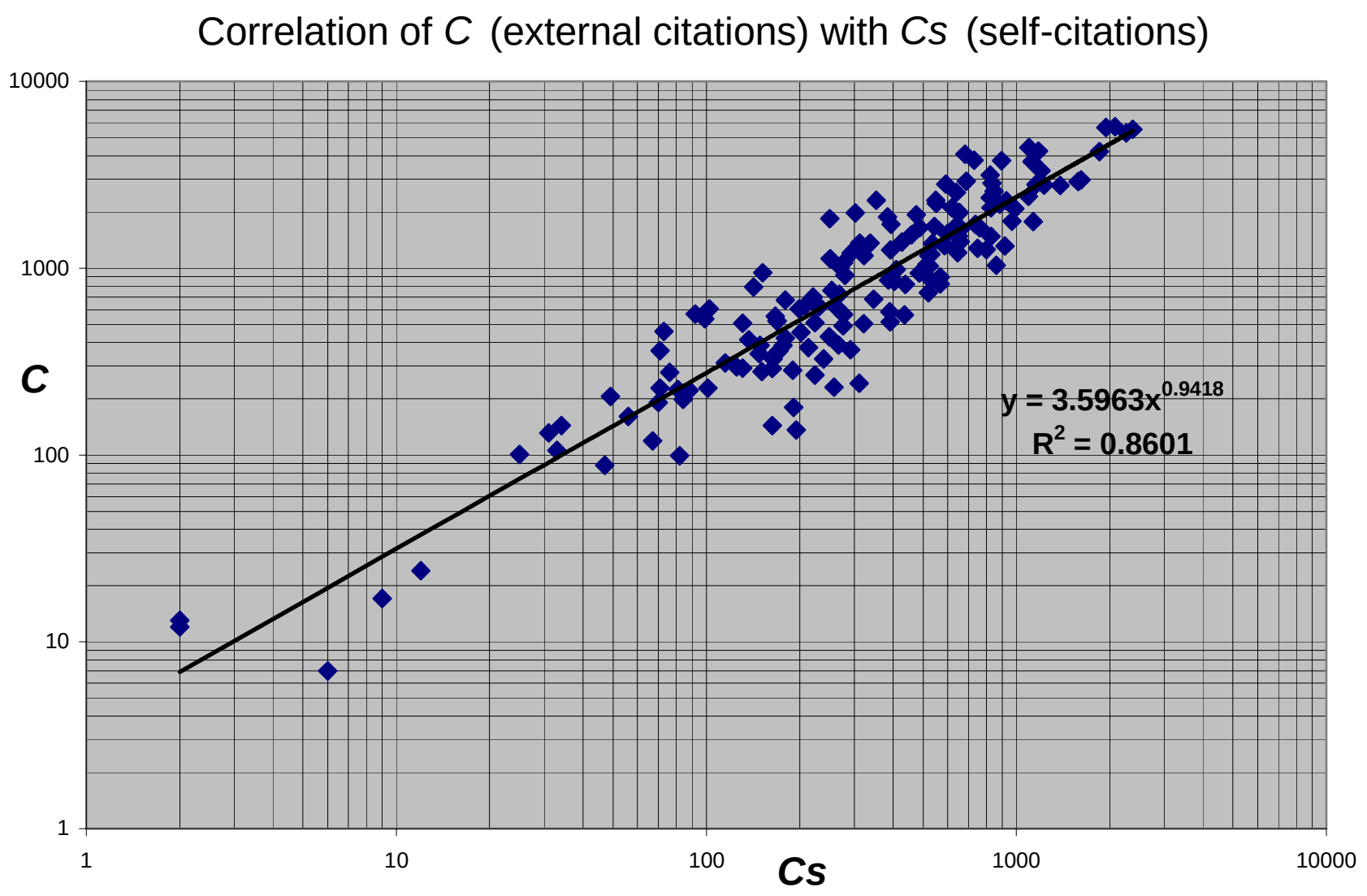

***Figure 3:*** *Correlation of the number of external citations **C** with the number of self-citations (**$C_S$**) received per chemistry group.*

This is a clear indication that size is a stronger determinant for self-citations than for external citations. In other words, ***P*** is a better predictor for the number of self-citations than for external citations. This phenomenon is also visible at both a higher aggregation level, namely entire universities (van Raan 2008a) and at a lower aggregation level, a set of individual scientists (Costas, Bordons, van Leeuwen, van Raan 2008). A further analysis is shown in Fig.4: the correlation of the number of self-citations with size for the *high* and the *low performance* groups (top-20% and bottom-20% of the ***CPP/FCSm*** distribution.

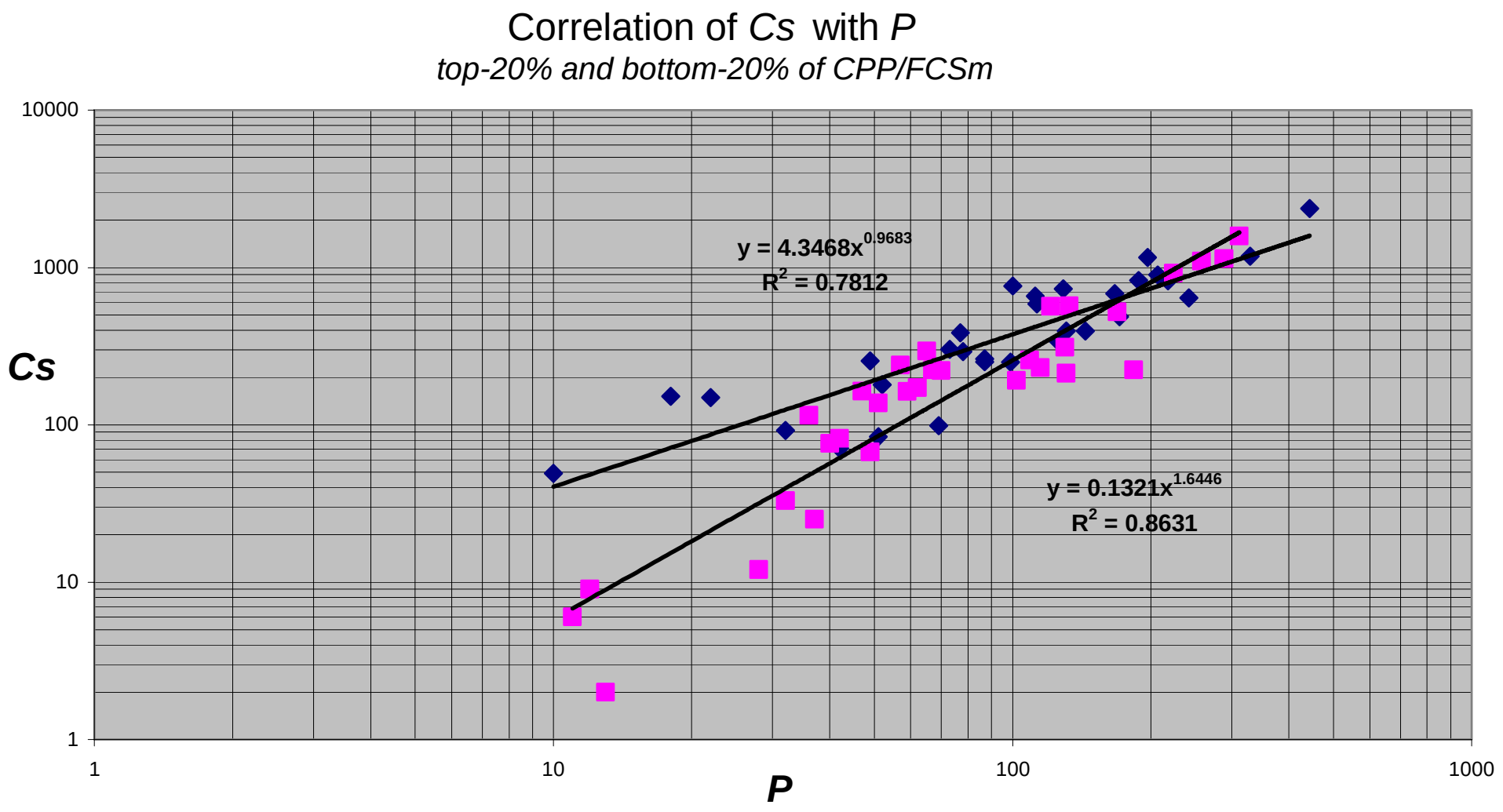

***Figure 4***: *Correlation of the number of self-citations (**$C_S$**) with the number of publications (**P**), for the top-20% groups (diamonds) and the bottom-20% groups (squares) of the **CPP/FCSm** distribution.*

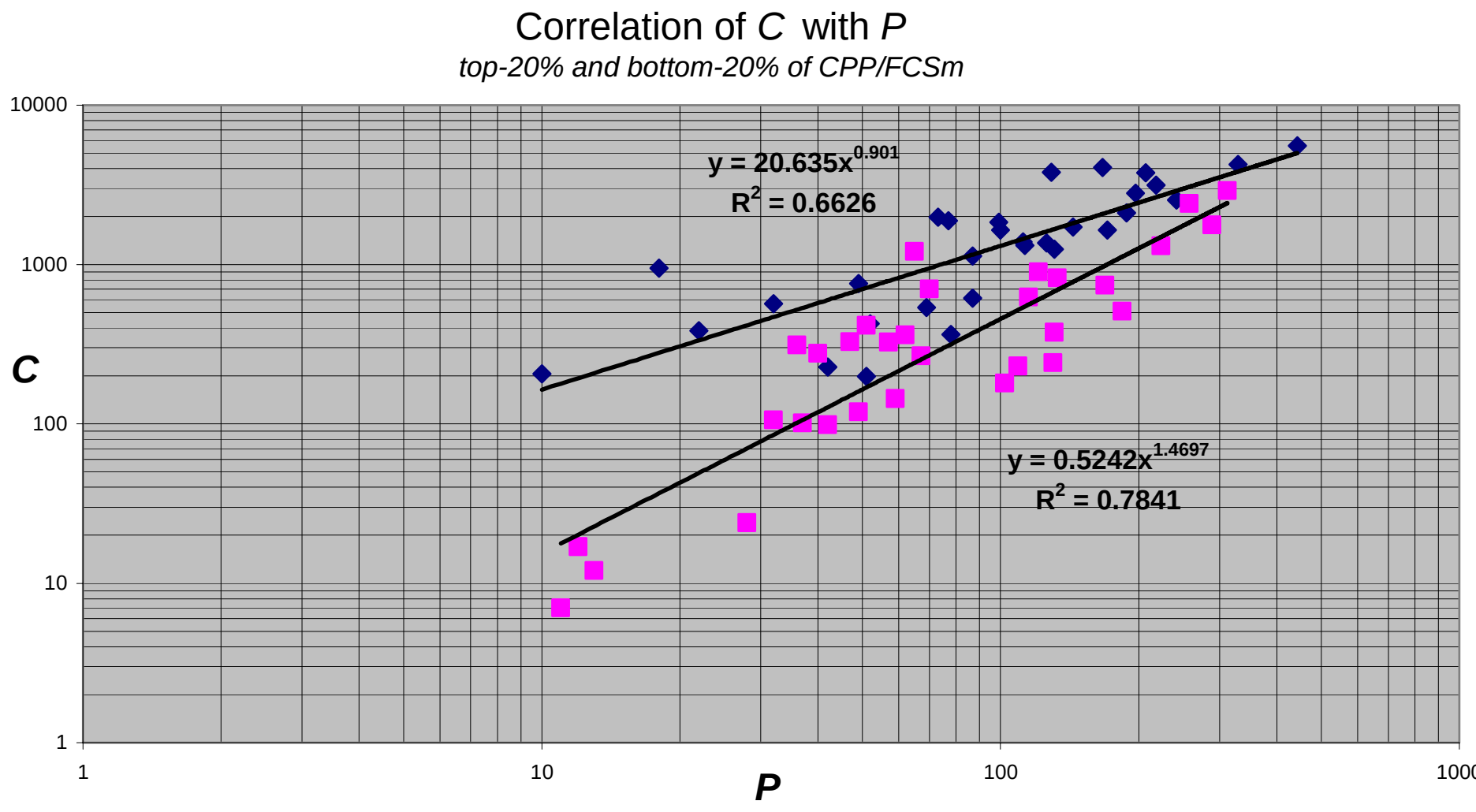

***Figure 5:*** *Correlation of the number of citations (**C**) with the number of publications (**P**), for the top-20% groups (diamonds), and for the bottom-20% groups (squares) of the **CPP/FCSm** distribution.*

We observe that top-performance groups have, as can be expected, more citations but also more self-citations (in an absolute sense!) than low performance groups. However, for top-performance group the number of self-citations increases more or less proportional with number of publications (power law exponent 0.97), whereas for the lower performance groups the number of self-citations increases in a nonlinear, cumulative way with the number of publications (power law exponent 1.64).

Comparison of Fig. 4 with Fig. 5 reveals that the power law exponents are higher for $C_s$ than for ***C***, and this is particularly the case for the lower performance groups. Thus, this is a further indication that size (***P***) enhances self-citations ($C_s$) more than external citations (***C***). We present an overview of the above results in Table 2. Remarkably, the value for ***P*** where the absolute number of citations and of self-citations are the same for the top and lower performance groups, is reached earlier (i.e., at smaller ***P***) for self-citations as compared to citations.

***Table 2****: Power law exponent* $\alpha$ *of the correlation of **C** and of* $C_s$ *with **P** for the all groups and for the top and lower performance groups*

| | ***all groups*** | ***groups with CPP/FCSm in*** | |
|---|---|---|---|
| | | *Top 20%* | *Bottom 20%* |
| ***C*** | 1.25 | 0.90 | 1.47 |
| $C_s$ | 1.33 | 0.97 | 1.64 |

Next, the correlation of $C_s$ with other bibliometric indicators such as field citation density and journal impact will be investigated. First we investigate the correlation of the fraction of self-citations ($C_s/C_i$) of the research groups with field citation density (***FCSm***). These results, presented in Fig. 6, show that there is no significant correlation.

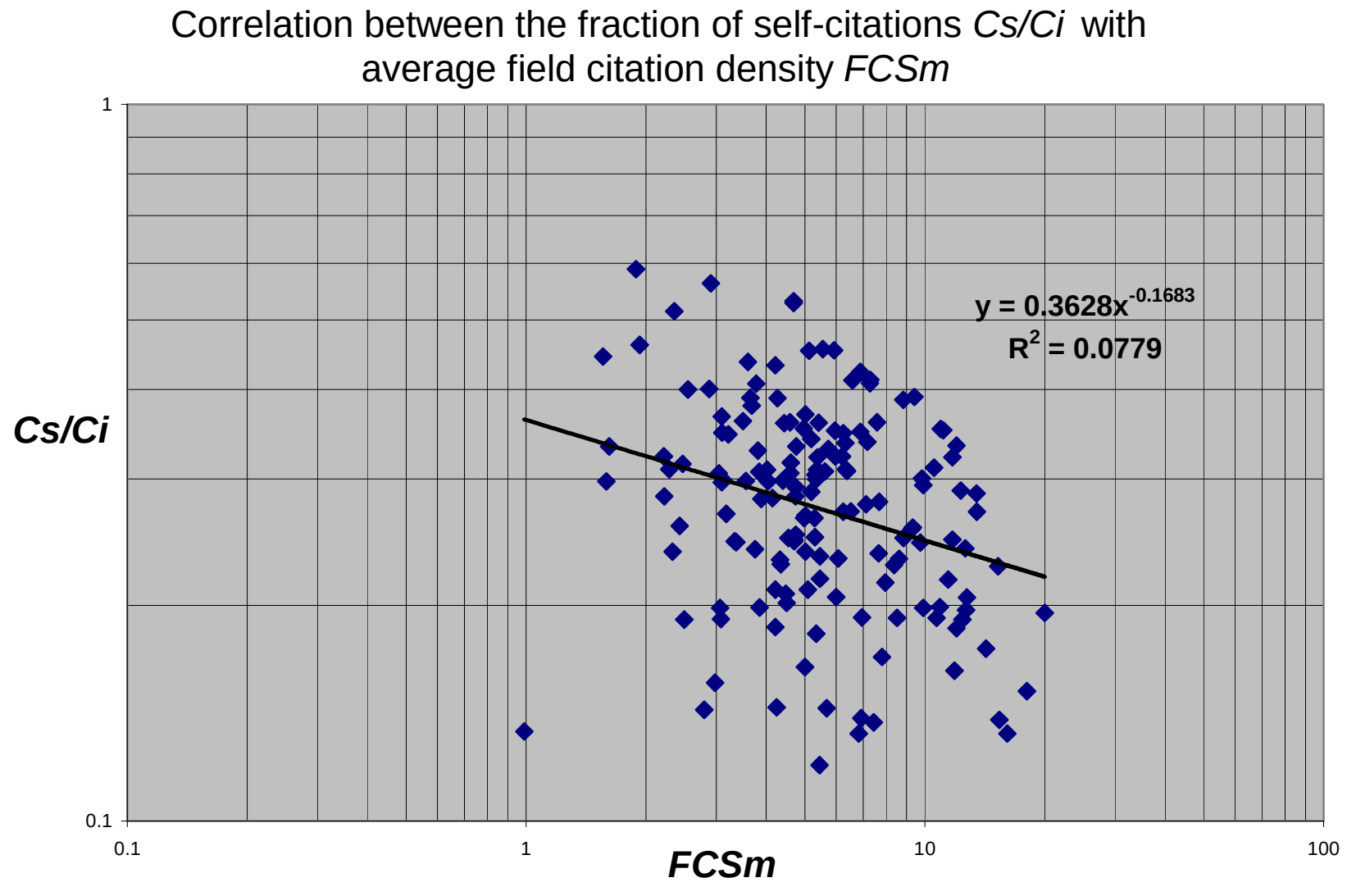

***Figure 6:*** *Correlation of the fraction of self-citations (**$C_s/C_i$**) with average field citation level (**FCSm**) for chemistry research groups. Seven groups with very small number of publications (**P** < 20) have been removed.*

For both the journal impact (***JCSm***) and the research performance (***CPP/FCSm***) a weak negative correlation with the fraction of self-citations is found, as shown in Figs. 7 and 8. Thus, the fraction of self-citations tends to decrease with journal impact and with performance.

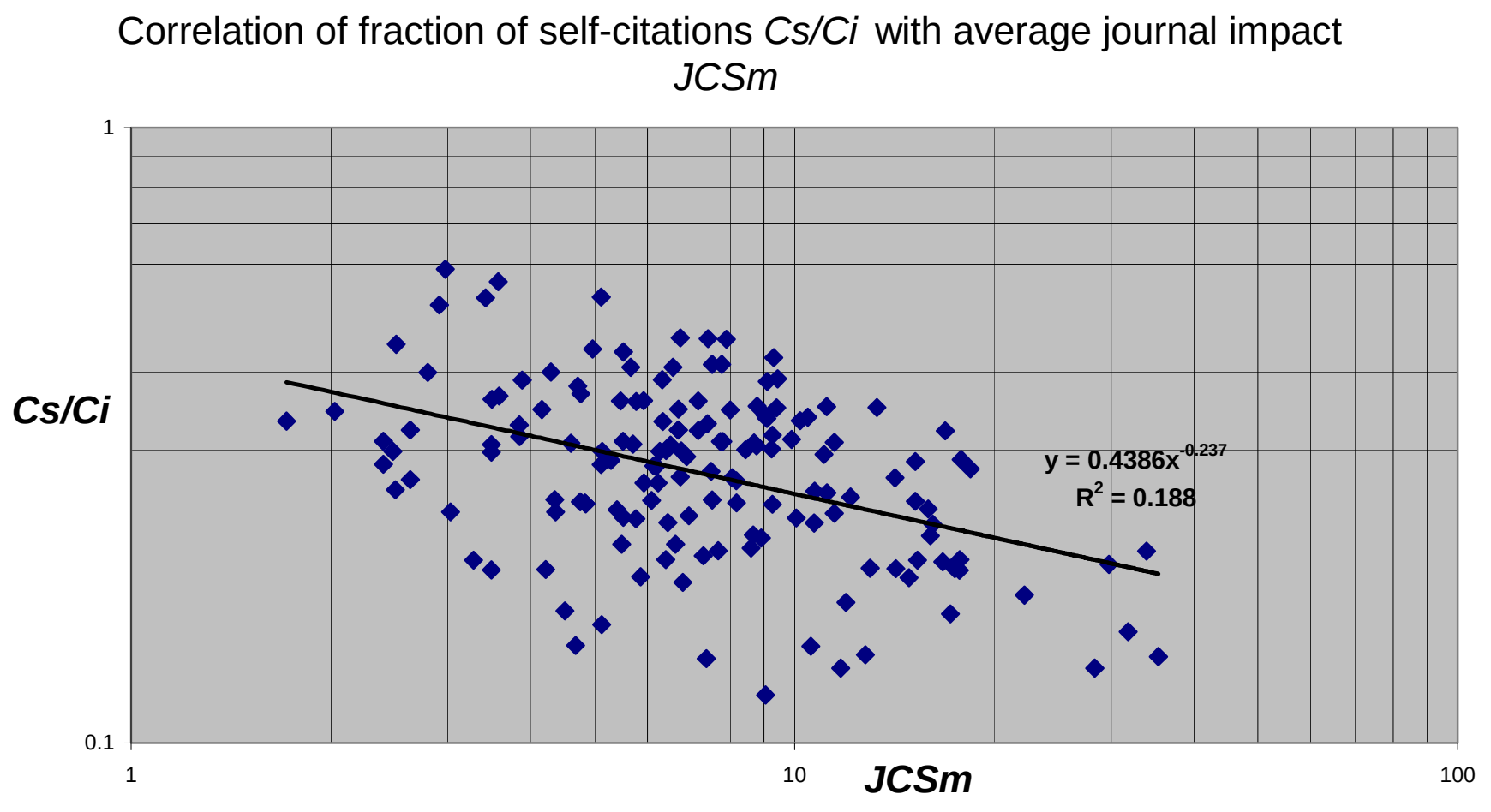

***Figure 7:*** *Correlation of the fraction of self-citations (**$C_s/C_i$**) with average journal impact (**JCSm**) for chemistry research groups. Seven groups with very small number of publications (**P** < 20) and/or very low journal impact (**JCSm** < 1.5) have been removed.*

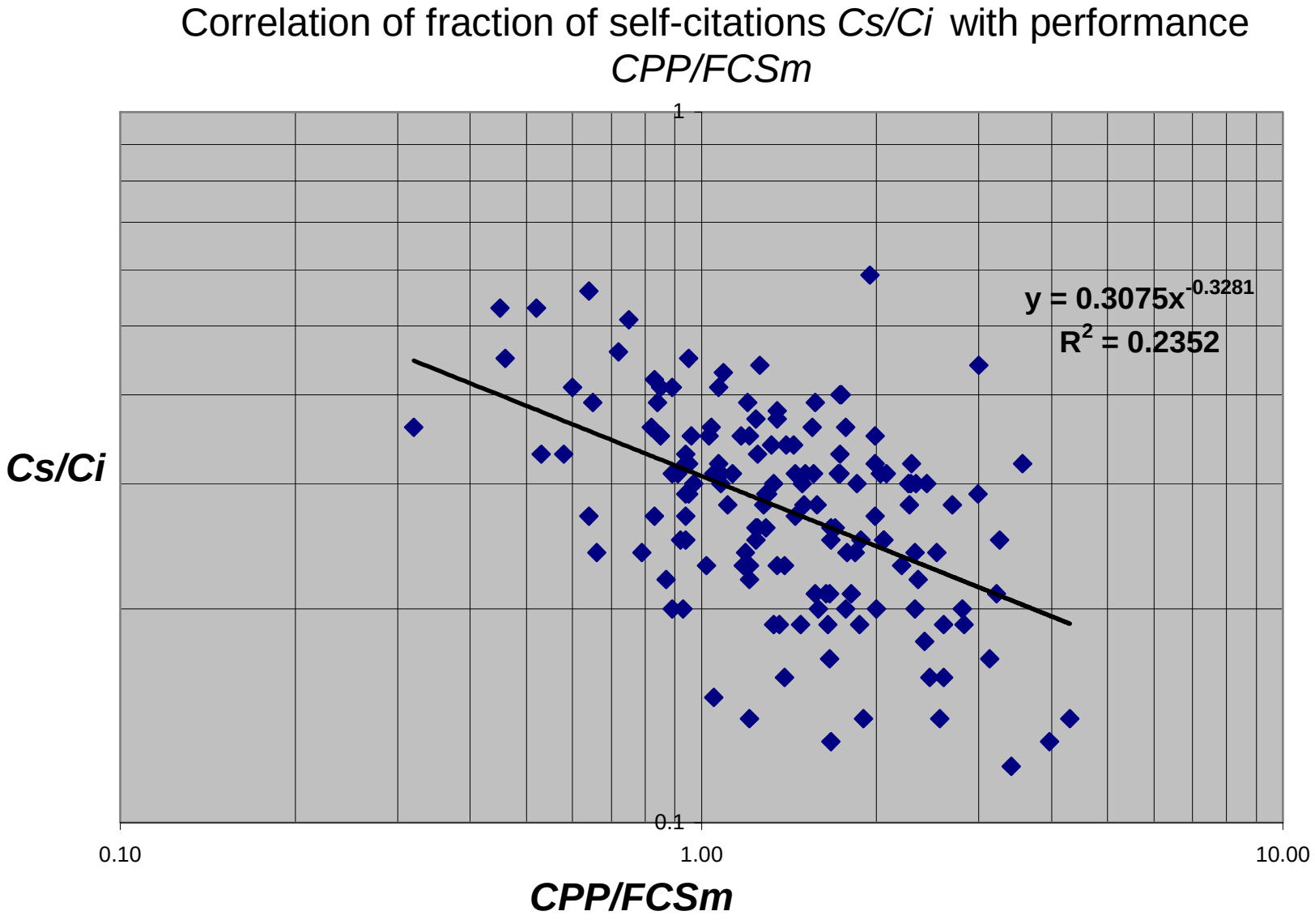

***Figure 8**: Correlation of the fraction of self-citations (**$C_s/C_i$**) with the research performance (**CPP/FCSm**) of the chemistry research groups.*

We find that size in terms of number of publications (***P***) hardly influences the correlation between self-citations and journal impact, but research performance (***CPP/FCSm***) does, see Fig. 9. This is the same figure as Fig. 7, but now only the groups in the top-20% and the bottom-20% of the performance distribution are selected. It can be observed that the top-performance groups (top-20% of the distribution) have relatively *less* self-citations than the lower performance groups (bottom-20% of the ***CPP/FCSm*** distribution), and this fraction is also decreasing more rapidly with journal impact.

The relatively strongest correlation of self-citations with basic bibliometric indicators is found for the average number of citations per publication (**CPP**) as shown in Fig. 10. The power law exponent of this correlation is -0.31. This finding comes rather close to the earlier reported square root dependence of the relative number of self-citations to the average number of citation per paper (Glänzel, Thijs and Schlemmer 2004; Glänzel, Debackere, Thijs and Schubert 2006). We stress however that our findings concern research groups whereas these earlier reported results concern a higher aggregation level, namely large institutions such as universities. In Section 4 we will show how the correlation between the number of self-citations and the average number of citations per publication follows directly from our model and the empirical results for the correlation of ***C*** and of ***$C_s$*** with ***P***.

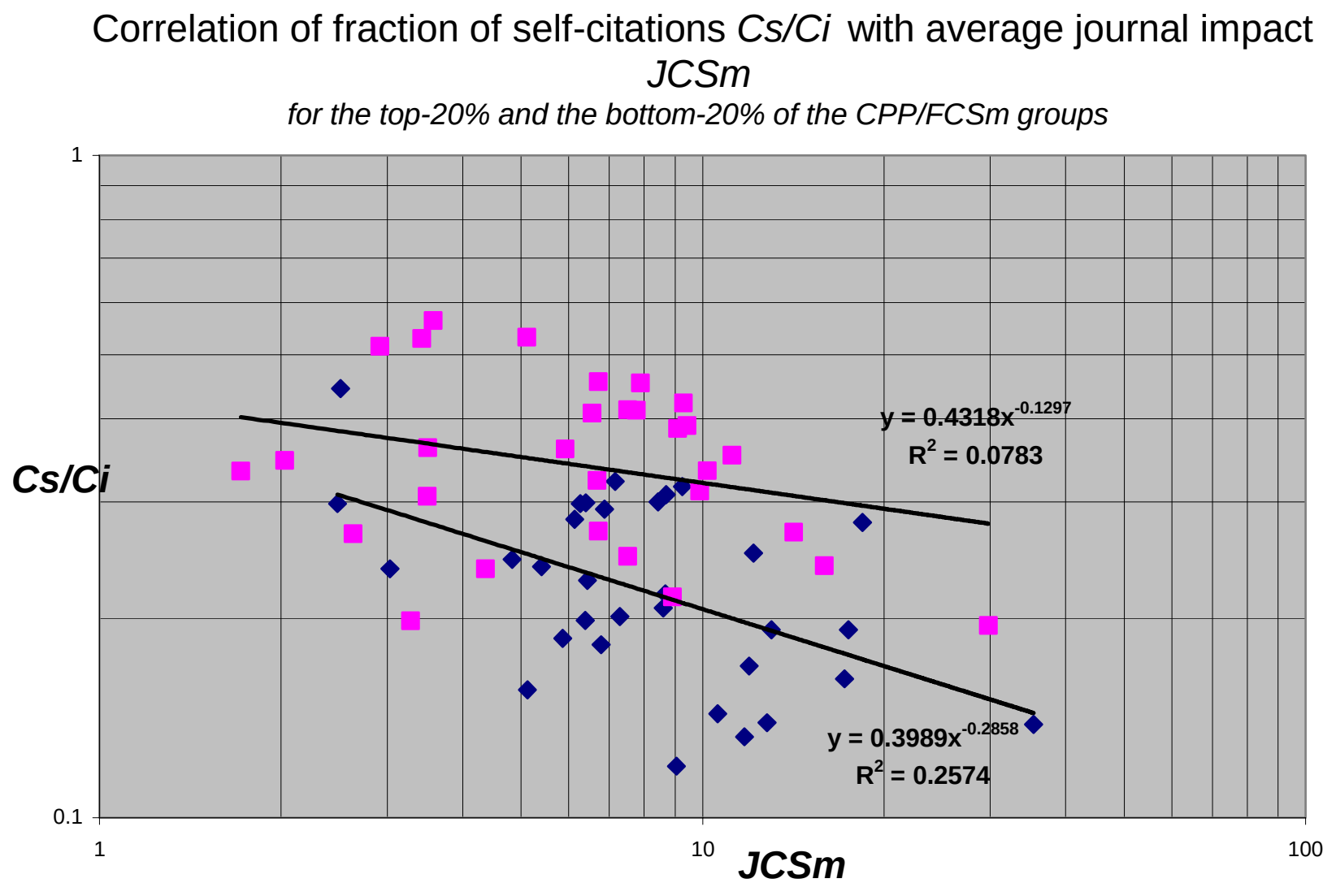

***Figure 9***: *Correlation of the fraction of self-citations (**$C_s/C_i$**) with average journal impact (**JCSm**) for the higher (diamonds) and the lower performance (squares) chemistry research groups, top-20% and bottom-20% of the **CPP/FCSm** distribution, respectively.*

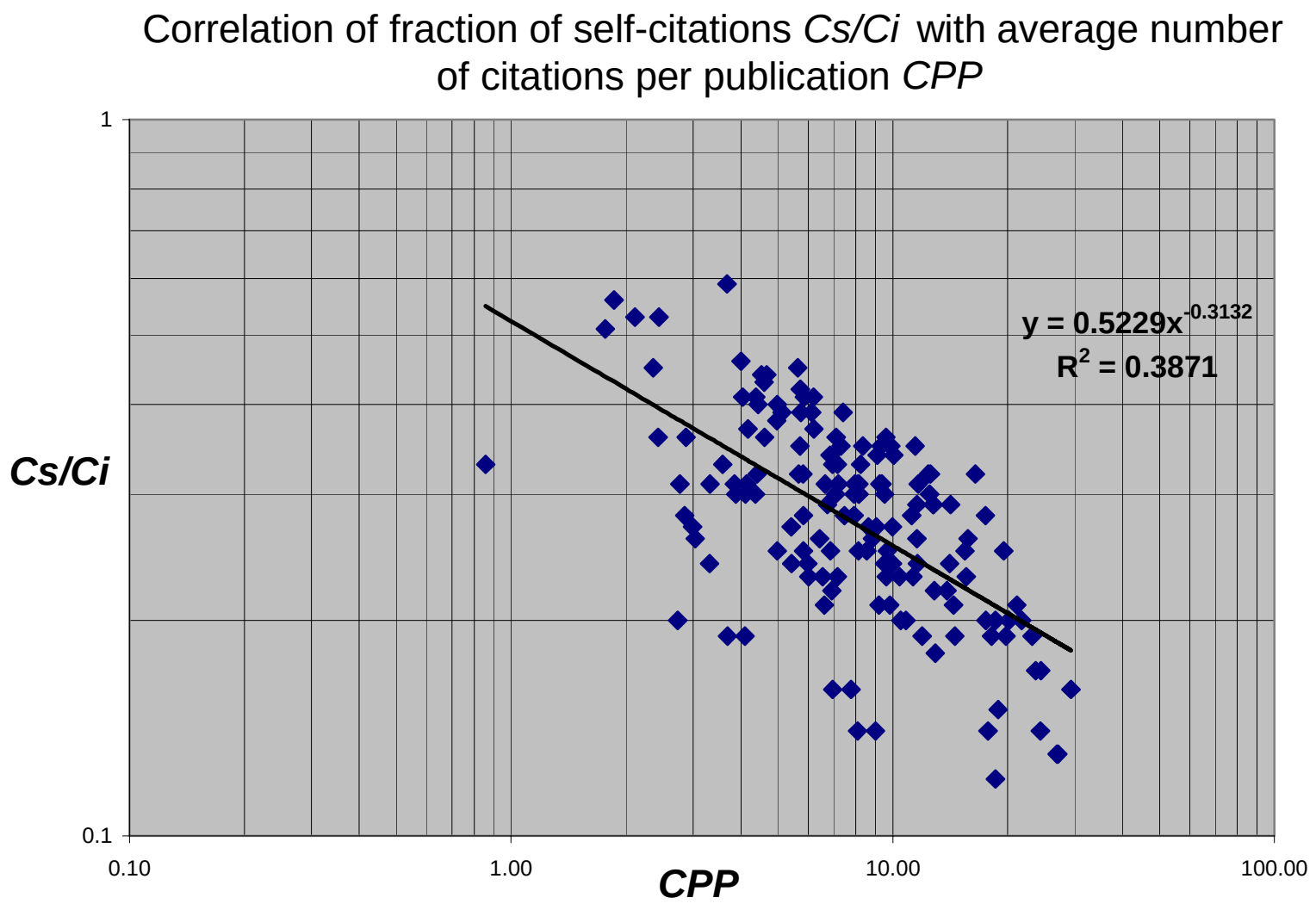

***Figure 10***: *Correlation of the fraction of self-citations (**$C_s/C_i$**) with average number of citations per publication (**CPP**).*

### *3.2 Influence of field citation density*

In a previous paper (van Raan 2008b) we studied the influence of *field citation density* on the impact of research groups. In this section we take a similar approach to investigate the influence of field citation density on self-citations. First we analyze the main overall characteristics by determining the correlation of the absolute number of self-citations ($C_S$) for the research groups with size in terms of absolute number of publications (***P***) for high and low field citation densities (top-20% and bottom-20% of the ***FCSm*** of research groups, respectively). The results are shown in Fig. 11.

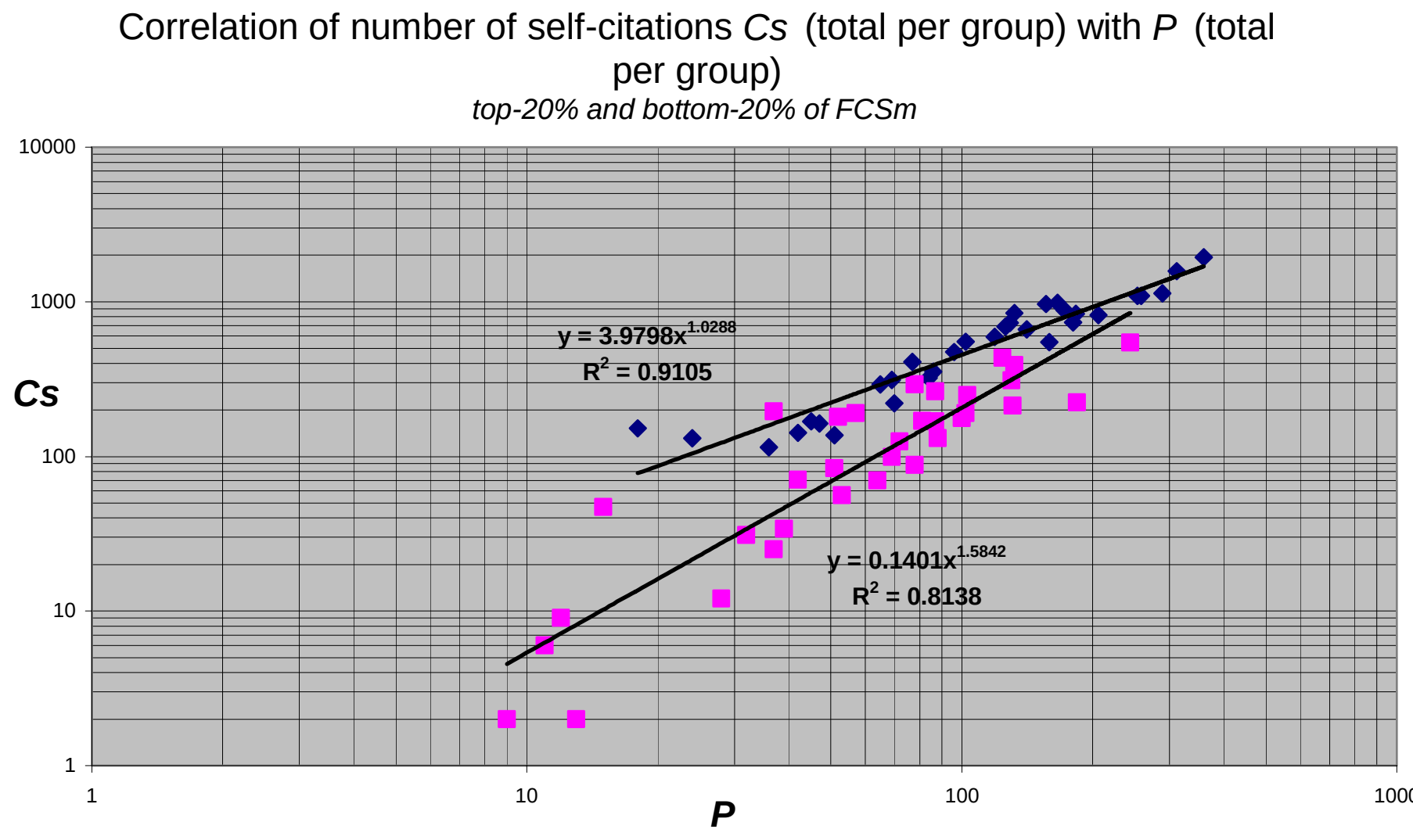

***Figure 11:*** *Correlation of the number of self-citations (*$C_S$*) with the number of publications (**P**), for groups in the top-50% (diamonds) and the bottom-50% (squares) of the field citation level (**FCSm**) distribution.*

We observe that research groups in fields with a relatively high citation density (top-20% of ***FCSm***) have more self-citations (as well as more citations) than groups in fields with a relatively low citation density (bottom-20% of ***FCSm***). But a remarkable observation is that the number of self-citations increases more or less proportionally with number of publications for research groups in fields with a relatively *high* citation density (power law exponent 1.03), and in a rather strongly cumulative way for research groups in fields with a relatively *low* citation density (power law exponent 1.58).

Next, the influence of field citation density on self-citations is investigated with distinction between top and lower performance groups. Fig. 12 presents the results for the *higher* performance groups (top-50% of the ***CPP/FCSm*** distribution) active in *high* versus *low* impact fields (top-20% and the bottom-20% of the ***FCSm*** distribution). It is found that only the high-performance groups in fields with low citation density (bottom-20% of ***FCSm***) have a size-dependent cumulative advantage (power law exponent 1.42) and therefore they reach for larger ***P*** the same total amount of self-citations as compared to the high-performance groups in high citation density fields.

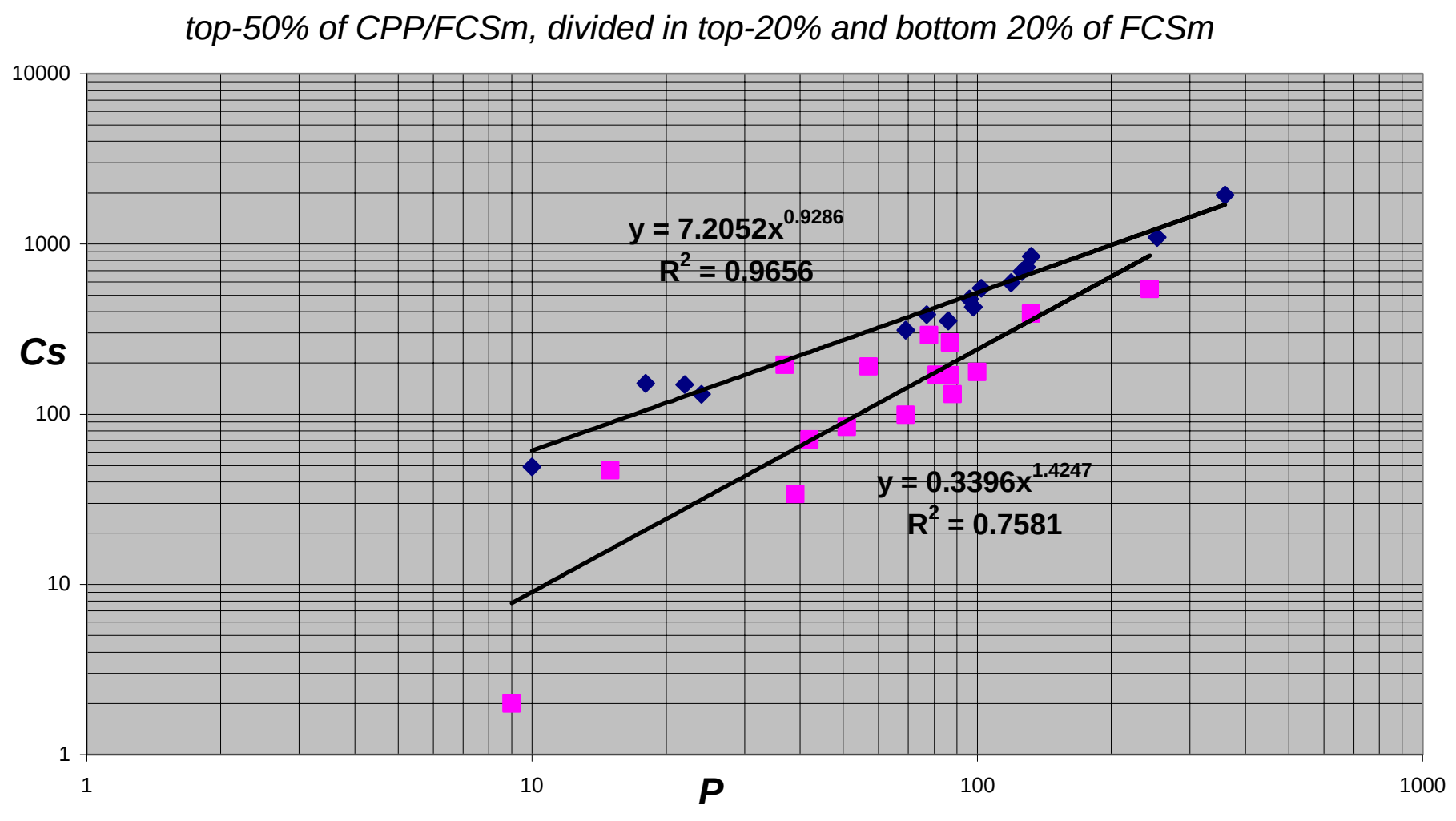

***Figure 12**: Correlation of the number of self-citations (**$C_S$**) with the number of publications (**P**) for the higher performance groups (top-50% of the **CPP/FCSm** distribution) divided in groups in the top-20% (diamonds) and in the bottom-20% (squares) of the average field citation level (**FCSm**).*

Do lower performance groups (bottom-50% of the ***CPP/FCSm*** distribution) behave differently in self-citation than higher performance groups with respect to field citation density? Fig. 13 answers this question. We see now in both cases a cumulative advantage with size, be it for the lower performance groups in fields with high citation density (top-20% of ***FCSm***) only modest (power law exponent 1.19) as compared to the strikingly large power law exponent of 1.70 for the groups in fields with high citation density (bottom-20% of ***FCSm***).

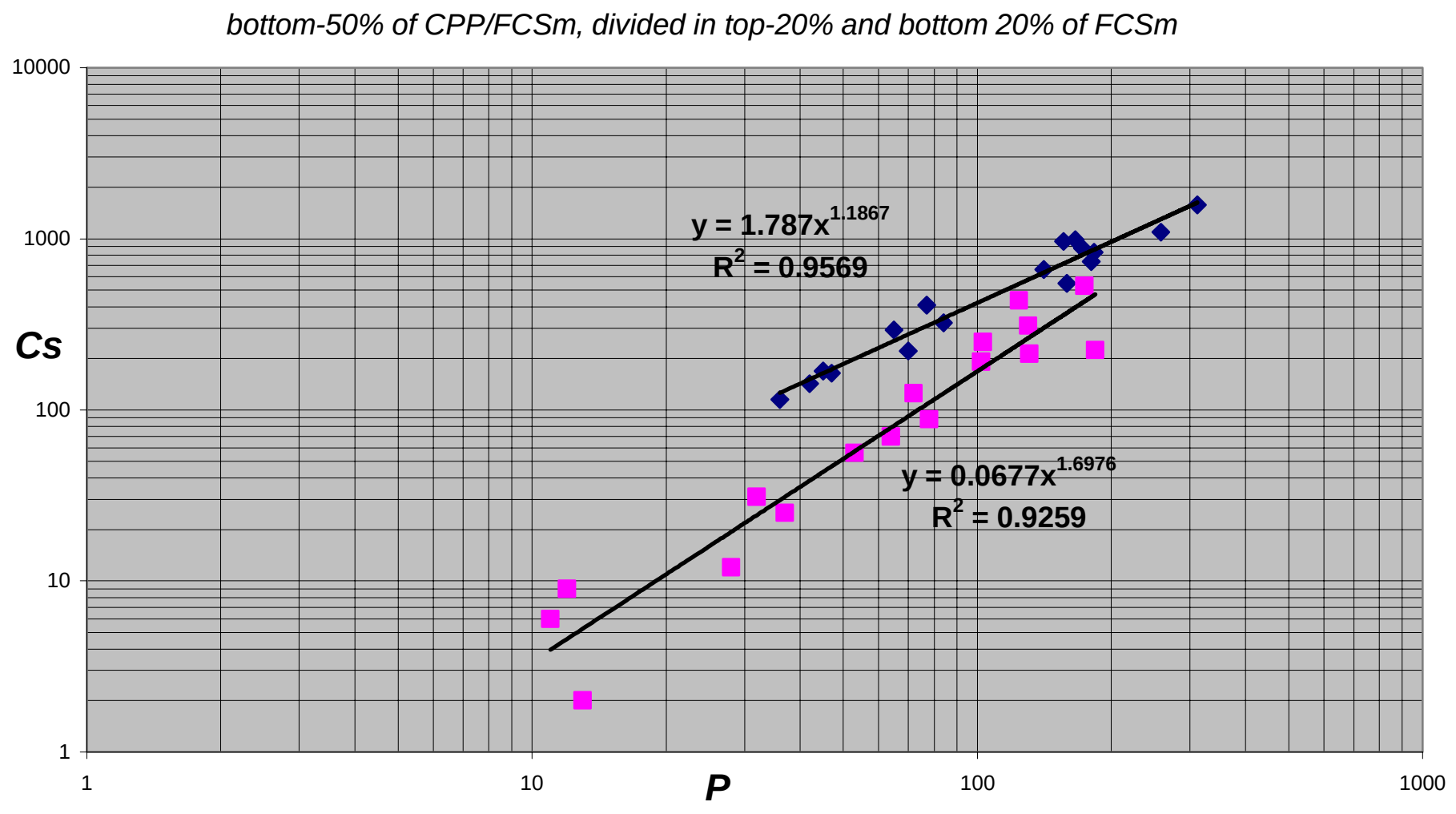

***Figure 13**: Correlation of the number of self-citations (**$C_S$**) with the number of publications (**P**) for the lower performance groups (bottom-50% of **CPP/FCSm** distribution) divided in groups in the top-20% (diamonds) and in the bottom-20% (squares) of the average field citation level (**FCSm**).*

We also studied the influence of *journal impact* on self-citation of research groups for both the ***JCSm*** as well as for the *field-normalized* journal impact indicator ***JCSm/FCSm***. The results are summarized in Table 3 where they are compared with the results for the field citation density as discussed in this section. Given the strong correlation of ***JCSm*** with ***FCSm*** at the level of research groups as shown in Fig. 14 we can expect that the results for these both indicators will not differ significantly. This is clearly shown in Table 3. Notice however that the relation between ***JCSm*** and ***FCSm*** is not a simple linear one, but as we see in Fig. 14 this relation is

$(\mathbf{JCSm}) \sim (\mathbf{FCSm})^{+1.06}$ .

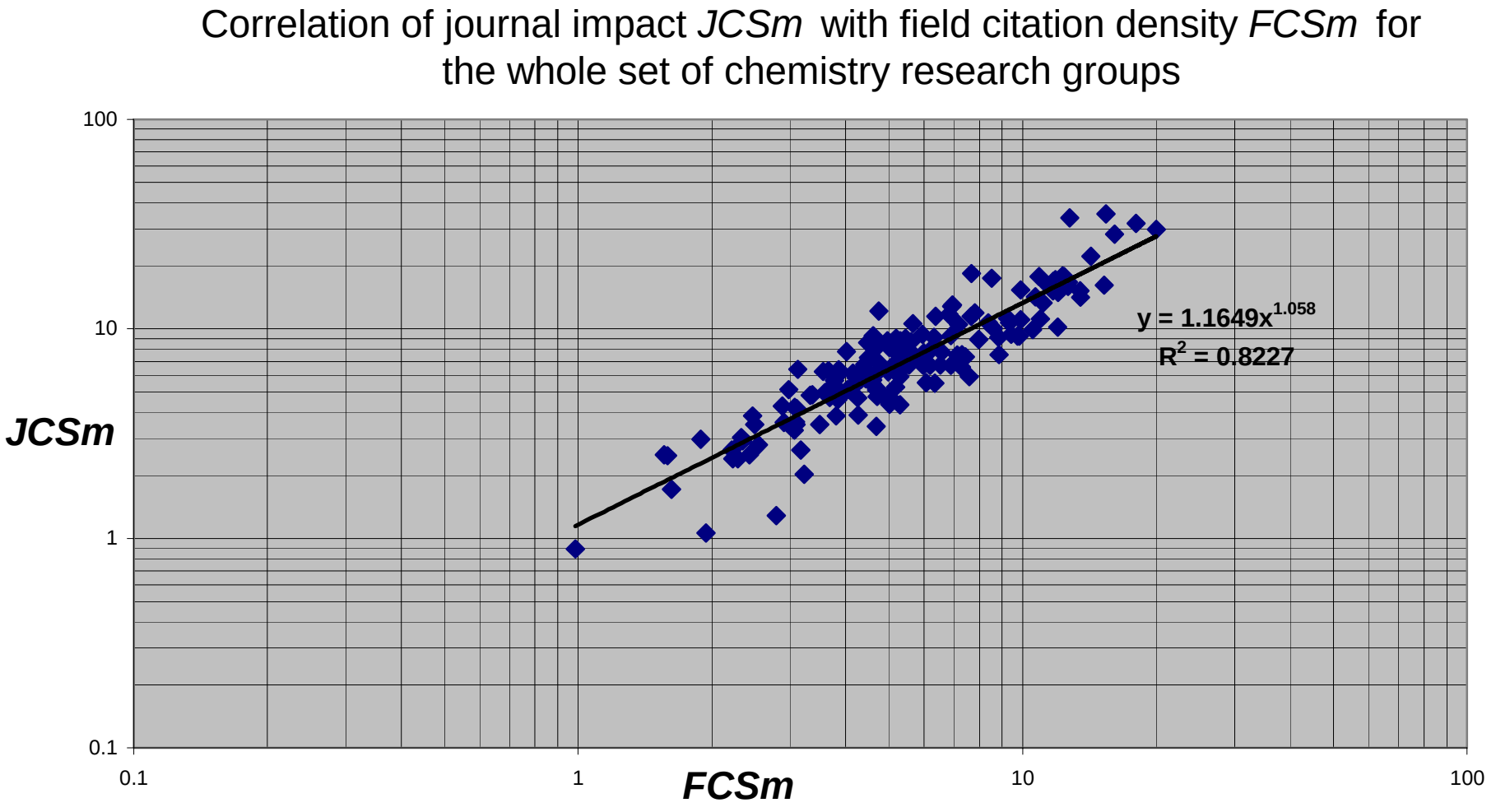

***Figure 14****: Correlation of the average journal impact (****JCSm****) with the average field citation density (****FCSm****) for all chemistry groups.*

***Table 3****: Power law exponent* ***α*** *of the correlation of* ***C*** *and of* ***Cs*** *with* ***P*** *with distinction between higher and lower performance groups (top-50% and bottom-50% of the* ***CPP/FCSm*** *distribution, respectively) as well as a distinction between high and low field citation density (top-20% and bottom-20% of the* ***FCSm*** *distribution, respectively), and high and low journal impact (top-20% and bottom-20% of the* ***JCSm*** *and the* ***JCSm/FCSm*** *distribution, respectively).*

| | | ***FCSm, JCSm, JCSm/FCSm*** | |
|---|---|---|---|
| | | *Top 20%* | *Bottom 20%* |
| ***CPP/FCSm*** | | | |
| | | | |
| *Top 50%* | ***C*** | 0.89, 0.94, 0.77 | 1.28, 1.28, 1.37 |
| | $\boldsymbol{C_s}$ | 0.93, 0.94, 0.89 | 1.42, 1.41, 1.59 |
| | | | |
| *Bottom 50%* | ***C*** | 0.97, 0.98, 1.34 | 1.43, 1.45, 1.62 |
| | $\boldsymbol{C_s}$ | 1.19, 1.19, 1.51 | 1.70, 1.67, 1.53 |

As discussed earlier, in most cases we find larger power law exponents for $\boldsymbol{C_s}$ than for ***C***, except for lower performance groups and low field-normalized journal impact.

In other words, for higher and particularly lower performance groups the size-dependence for self-citations is generally stronger than for citations, but for the higher performance groups this is only the case for the lower impact fields and journals. Furthermore, the table also shows that this enhancement by size of both self- and non-self citations is *largest for lower field citation density and particularly for the low impact journals*. Thus, lower impact journals 'make size work', also for higher performance groups.

To distinguish in more detail between the influence of the field citation density and of average journal impact, we first present in Fig. 15 the results for the size-dependence of self-citations for the *higher* field citation density groups (top-50% of the ***FCSm*** distribution) publishing in *higher* versus *lower* impact journals (top-20% and bottom-20% of the ***JCSm*** distribution). The striking role of lower impact journals is confirmed again: only for these lower impact journals the number of self-citations increases with size with considerable cumulative advantage (power law exponent 1.31). Fig. 16 shows a similar behavior for low performance groups (bottom-20% of the ***CPP/FCSm*** distribution).

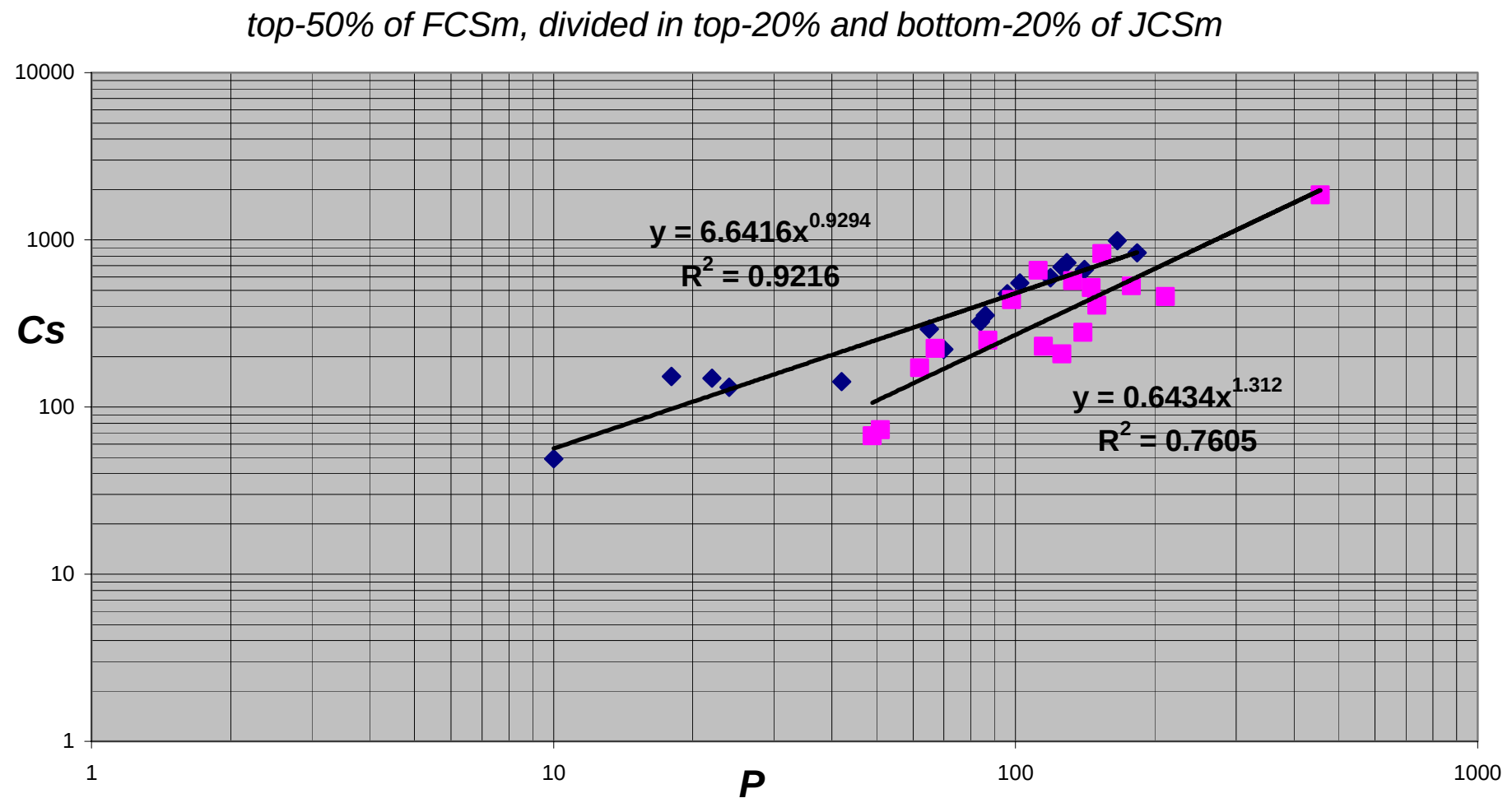

***Figure 15:*** *Correlation of the number of self-citations (**$C_s$**) with the number of publications (**P**) for research groups in fields with a higher citation density (top-50% of **FCSm**) divided in groups publishing in the top-20% (diamonds) and in the bottom-20% (squares) of the **JCSm** distribution.*

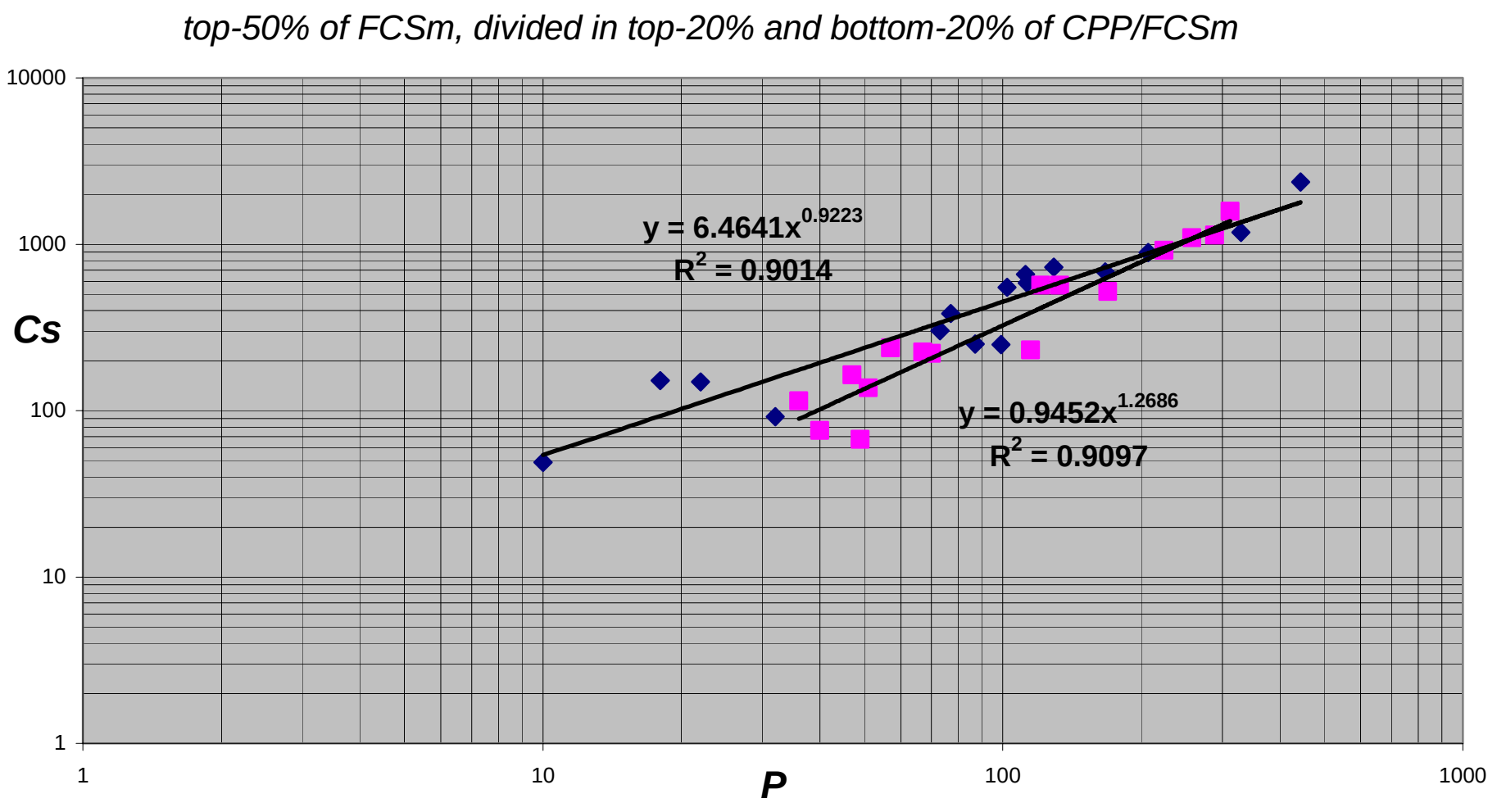

***Figure 16:*** *Correlation of the number of self-citations (****$C_s$****) with the number of publications (****P****) for research groups in fields with a higher citation density (top-50% of* ***FCSm****) divided in higher and lower performance groups: top-20% (diamonds) and in the bottom-20% (squares) of the* ***CPP/FCSm*** *distribution.*

In Fig. 17 we show the size-dependence of self-citations for the *lower* field citation density groups (bottom-50% of the **FCSm** distribution) publishing in *higher* versus *lower* impact journals (top-20% and bottom-20% of the **JCSm** distribution). The findings are quite striking: the number of self-citations for groups that are both in fields with lower citation density *and* publishing in the low impact journals increase almost with the square (power law exponent 1.83) of the number of publications. Within our set of observations this is the most extreme size-dependence.

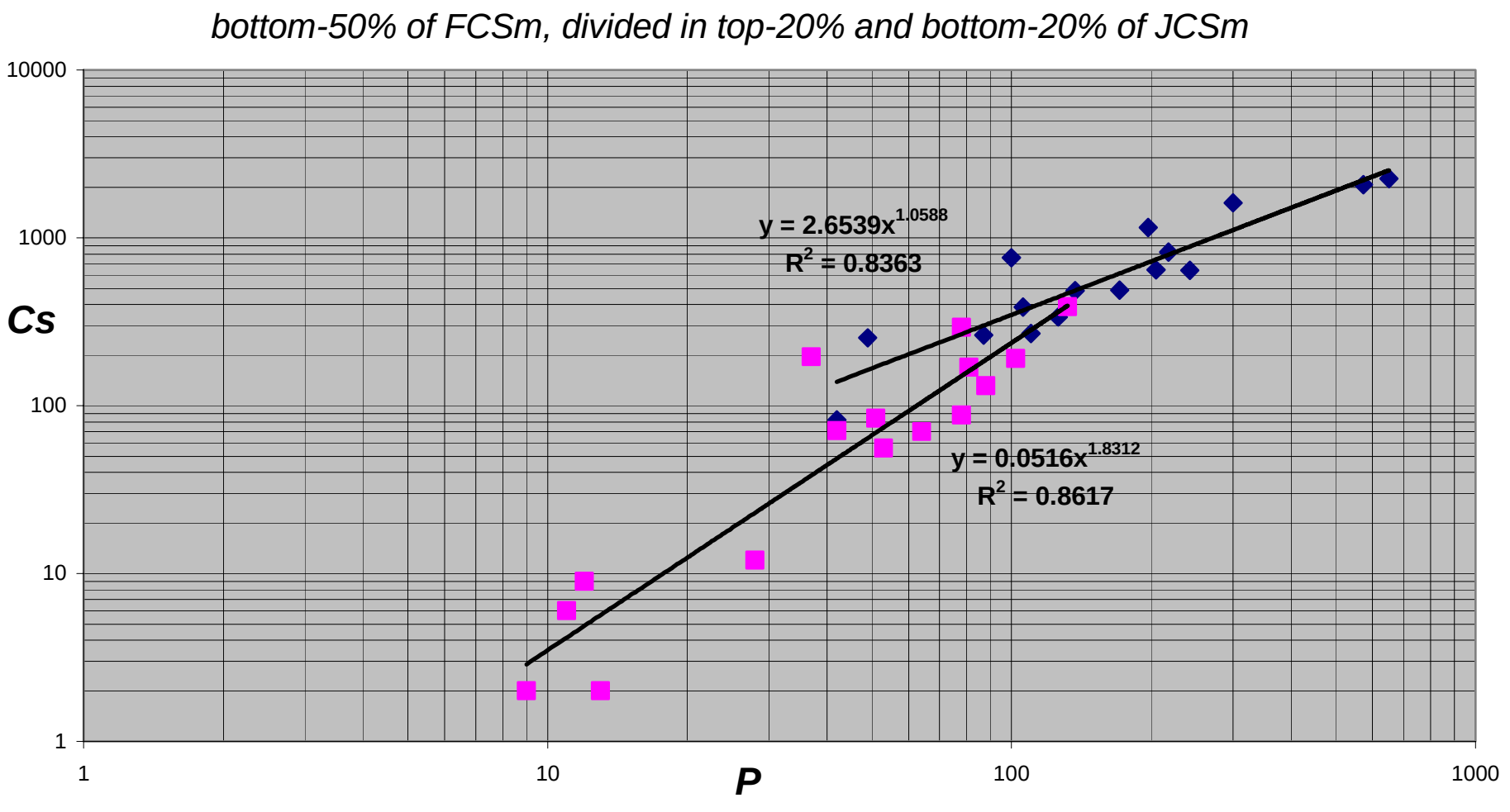

***Figure 17:*** *Correlation of the number of self-citations (****$C_s$****) with the number of publications (****P****) for research groups in fields with a lower citation density (bottom-50% of* ***FCSm****) divided in groups publishing in the top-20% (diamonds) and in the bottom-20% (squares) of the* ***JCSm*** *distribution.*

Fig. 18 shows a similar but less extreme behavior for low performance groups (bottom-20% of the ***CPP/FCSm*** distribution).

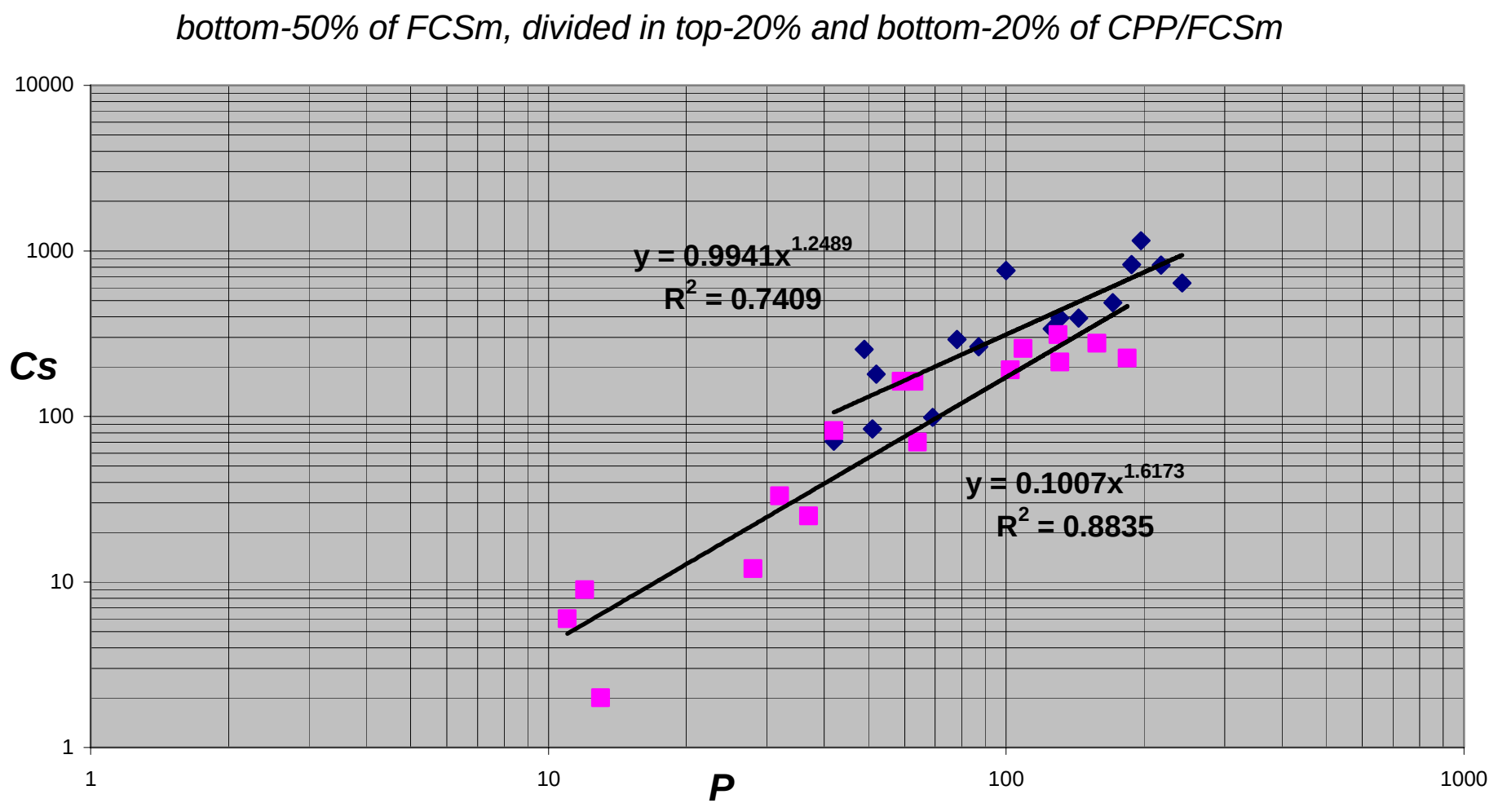

***Figure 18:*** *Correlation of the number of self-citations (**$C_s$**) with the number of publications (**P**) for research groups in fields with a lower citation density (bottom-50% of **FCSm**) divided in high (top-20%, diamonds) and lower performance groups (bottom-20%, squares, of the **CPP/FCSm** distribution.*

We also studied the influence of the field-normalized ***JCSm/FCSm*** indicator on the size-dependence of self-citation. Together with the above discussed findings the results are summarized in Table 4.

***Table 4****: Power law exponent **α** of the correlation of **C** and of **$C_s$** with **P** with distinction between groups in fields with higher and lower citation density (top50% and bottom-50% of the **FCSm** distribution, respectively) as well as a distinction between high and low performance (top-20% and bottom-20% of the **CPP/FCSm** distribution, respectively), and high and low journal impact (top-20% and bottom-20% of the **JCSm** and the **JCSm/FCSm** distribution, respectively).*

| | | ***CPP/FCSm, JCSm, JCSm/FCSm*** | |
|---|---|---|---|
| | | *Top 20%* | *Bottom 20%* |
| ***FCSm*** | | | |
| | | | |
| *Top 50%* | ***C*** | 0.82, 0.85, 0.84 | 1.08, 1.26, 1.05 |
| | ***$C_s$*** | 0.92, 0.86, 0.93 | 1.27, 1.27, 1.31 |
| | | | |
| *Bottom 50%* | ***C*** | 1,40, 1.06, 1.17 | 1.39, 1.57, 1.49 |
| | ***$C_s$*** | 1.25, 0.99, 1.06 | 1.62, 1.83, 1.68 |

## 4. Discussion of the results in the framework of the model

In previous papers statistical properties of bibliometric indicators at the aggregation level of research groups were discussed (van Raan 2006b) and in particular the influence of field-specific citation characteristics (van Raan 2008b). It was demonstrated that lower performance groups have a larger size-dependent cumulative advantage for receiving citations than top-performance groups. Furthermore, regardless of performance, larger groups have less not-cited publications. Particularly for the lower performance groups the fraction of not-cited publications decreases considerably with size.

In this article we show that for self-citations similar size-dependent scaling rules apply as for citations but generally the power law exponents are higher for self-citations as compared to citations. Furthermore we find that the fraction of self-citations is smaller for the higher performance groups and this fraction decreases more rapidly with increasing journal impact than for lower performance groups. An interesting novel finding is that the variance in the correlation of the number of self-citations with size is considerably less than the variance for external citations. This is a clear indication that size is a stronger determinant for self-citations than for external citations.

The main properties of these size-dependent phenomena can be explained with the model (Mechanism A) in which self-citation acts as a promotion mechanism for external citations (van Raan 2006b; Fowler and Aksnes 2007). The idea behind mechanism A is that advantage by size works by a process in which the number of not-cited publications is diminished, and that this mechanism is particularly effective for the lower performance groups. Thus, the larger the number of publications in a group, the more those publications are 'promoted' which otherwise would have remained uncited. Most probably this works by initial citation of these 'stay behind' publications in other more cited publications of the group. Then authors in other groups are stimulated to take notice of these stay behind publications and they eventually decide to cite them. Consequently, the mechanism starts with within-group self-citation, and subsequently spreads.

The Mechanism A model can be described mathematically as follows. If size reinforces self-citation, the relative increase of self-citations $(\Delta C_s)/C_s$ will be *larger* than the relative increase of the number of publications $(\Delta P)/P$ and thus it can be written in first approximation by

$$(\Delta C_s)/C_s = \alpha(\Delta P)/P \qquad \text{(Eq. 1a)}$$

which yields with $a$ as integration constant

$$C_s = aP^{\alpha} \qquad \text{(Eq. 1b)}$$

Furthermore, if self-citation promotes external citation the relative increase of citations $(\Delta C)/C$ will be larger than the relative increase of the number of self-citations $(\Delta C_s)/C_s$ so that in first approximation

$$(\Delta C)/C = \beta(\Delta C_s)/C_s \qquad \text{(Eq. 2a)}$$

which yields with $\boldsymbol{b}$ as integration constant

$$C = bC_s^{\beta} \qquad \text{(Eq. 2b)}$$

By combining Eqs. 1b and 2b we find

$$C = baP^{\alpha\beta} \qquad \text{(Eq. 2c)}$$

From Eq. 2b follows with $b' = b^{1/\beta}$ that $C_s = (1/b')\, C^{(1/\beta)}$ and with help of Eq. 2c we find by taking $(1/b')(ba)^{[(1/\beta)-1]} = b''$

$$C_s/C = b''\, P^{\alpha\beta[(1/\beta)-1]} = b''\, P^{\alpha(1-\beta)} \qquad \text{(Eq. 2d)}$$

From Eq. 2c follows $C/P = baP^{(\alpha\beta-1)}$ so that with $a' = (ba)^{1/(\alpha\beta-1)}$

$$(C/P)^{1/(\alpha\beta-1)} = a'P \qquad \text{(Eq. 3a)}$$

Combining Eq. 3a with Eq. 2d we find by taking $B = b''/a'$

$$C_s/C = B\,(C/P)^{\alpha(1-\beta)/(\alpha\beta-1)} \qquad \text{(Eq. 3b)}$$

From the empirical results it is found that $\alpha = 1.33$ (see Fig. 2) and $\beta = 0.94$ (see Fig. 3). Inserting these empirical values yields

$$\alpha(1-\beta)/(\alpha\beta-1) = 0.32$$

and thus we find

$$C_s/C \sim (C/P)^{0.32}$$

which is confirmed very well by Fig. 10 where we showed the correlation between $C_s/C_i$ and the indicator '***CPP***', which is $C/P$. Because $C_i = (C+C_s)^{1.02}$, $C$ is in good approximation equal to the total amount of citations $C_i$ (i.e., including self citations). Thus it is explained empirically that the relative number of self-citations is related to the number of citations per publications with a power law exponent around 0.3.

In previous work (van Raan 2006b) we found that *only* lower performance groups have a significant size-dependent cumulative advantage for the absolute number of received (external) citations ($C$), for higher performance groups this cumulative advantage is found only for groups in the lower impact journals and fields. A quite intriguing finding is that the cumulative size-dependent advantage is largest for lower performance groups publishing in journals and fields with relatively low impact. In the remarkable case where lower performance groups publish in journals and fields with relatively high impact, hardly or no cumulative size-dependent advantage is found.

In this study we find very similar results for the statistics of self-citation: again, *only* lower performance groups have a size-dependent cumulative advantage for the absolute number of self-citations of research groups ($C_s$), for higher performance groups this cumulative advantage is found only for groups in the lower impact journals and fields. The largest size–dependent cumulative advantage is found for research groups in fields with a lower citation density *and* publishing in low impact journals. The important difference between the scaling of citations and of self-citations however is that the power law exponents are *higher* for $C_s$ as compared to $C$, and this is particularly the case for the lower performance groups in both the top-20% and the bottom-20% of the average journal impact of the group (***JCSm***). We also find that the fraction of self-citations tend to decrease with journal impact and with performance.

From our findings in this study we conclude that, generally, size ($P$) enhances self-citations ($C_s$) more than (external) citations ($C$), and this finding is further supported by the observation that $P$ is a better predictor for $Cs$ than for $C$. The model discussed in this article (Mechanism A) implies that external citations are enhanced by self-citations, so that we have the 'chain reaction': larger size leads to more self-citations which on their turn lead to more external citations. This mechanism is strongest for the lower impact journals, they 'make size work', also for higher performance groups. In other words, lower impact journals enable research groups more than higher impact journals to 'advertise' their other work by means of self-citations.

**_Acknowledgements_**

The author would like to thank his CWTS colleague Thed van Leeuwen for the data collection, data analysis and calculation of the bibliometric indicators for the two sets of research groups.